\documentclass[12pt]{article}
\usepackage[T1]{fontenc}
\usepackage[utf8]{inputenc}
\usepackage[margin=1in]{geometry}
\usepackage[affil-it]{authblk} 
\usepackage{etoolbox}
\usepackage{lmodern}
\usepackage{cite}
\usepackage{changes}
\usepackage{microtype}
\usepackage{booktabs}
\usepackage{amssymb}
\usepackage{float}
\DeclareUnicodeCharacter{030C}{\v{c}}
\usepackage{nameref}
\title{Wafer-Scale Integration of Freestanding Photonic Devices with Color Centers in Silicon Carbide}

\def\correspondingauthor{\footnote{Corresponding author: smajety@ucdavis.edu}}

\author[1] {Sridhar Majety \correspondingauthor{}}
\author[1, 2] {Victoria A. Norman}
\author[1] {Pranta Saha}
\author[1, 2] {Alex H. Rubin}
\author[3] {Scott Dhuey}
\author[1] {Marina Radulaski}

\affil[1]{Department of Electrical and Computer Engineering, University of California, Davis, CA 95616, USA}
\affil[2]{Department of Physics and Astronomy, University of California, Davis, CA 95616, USA}
\affil[3]{The Molecular Foundry, Lawrence Berkeley National Laboratory, Berkeley, CA 94720, USA}

\date{\vspace{-2em}}

\usepackage{graphicx}
\usepackage{amsmath}
\usepackage{braket}
\usepackage{titlesec}
\usepackage{siunitx}
\usepackage[version=4]{mhchem}
\usepackage{multirow}
\begin{document}

\maketitle
\vspace{-0.8cm}
\begin{abstract}

Color center platforms have been at the forefront of quantum nanophotonics for applications in quantum networking, computing, and sensing. However, large-scale deployment of this technology has been stifled by a lack of ability to integrate photonic devices at scale while maintaining the properties of quantum emitters. We address this challenge in silicon carbide, which has both commercially available wafer-scale substrates and is a host to color centers with desirable optical and spin properties. Using ion beam etching at an angle, we develop a 5-inch wafer process for the fabrication of 
triangular cross-section photonic devices in bulk 4H-SiC. The developed process has a variability in etch rate and etch angle of 5.4\% and 2.9\%, respectively. Furthermore, the integrated color centers maintain their optical properties after the etch, thus achieving the nanofabrication goal of wafer-scale nanofabrication in quantum-grade silicon carbide.  
\end{abstract}
\section{Introduction}
Nanophotonics has gained prominence over the last few decades for applications in classical fields of communications, nonlinear optics, sensing and displays, as well as the quantum field of information processing (QIP). Solid-state emitters with optically addressable spin states have been the frontrunners for scalable QIP implementation due to their capability of creating entanglement between electron spin and photon, and subsequently, the emitted photon can be used to distribute entanglement over long distances. When such an emitter, with the ability to generate indistinguishable spin-entangled photons, is integrated into a nanophotonic structure, it can reach the efficiencies in emission and collection of photons \cite{norman2021novel, majety2022quantum}, necessary to implement QIP protocols. Earlier efforts in this direction using quantum dots and the nitrogen vacancy color center in diamond demonstrated groundbreaking effects of quantum nanophotonics but struggled to reach scalability due to either the lack of scalable nanofabrication or the lack of spectral uniformity among emitters. Silicon carbide (SiC) promises to bridge these gaps. SiC is a well established host of color centers that combine long spin coherence times \cite{anderson2021five}, excellent brightness \cite{castelletto2020silicon}, room-temperature spin manipulation \cite{wang2020NVSiC}, ample nuclear spins \cite{bourassa2020entanglement}, ultra-narrow inhomogeneous broadening \cite{cilibrizzi2023ultra}, and emission wavelengths in the near-infrared and telecom bands \cite{majety2022quantum}. Most importantly, SiC is available as a wafer-scale substrate, and is compatible for nanofabrication in CMOS tools, making it a suitable platform for large-scale quantum nanophotonic hardware.

Suspended photonics offer the best optical confinement due to their highest refractive index contrast with the surrounding media, air (\textit{n} = 1). Such suspended photonics are mainly fabricated through heteroepitaxial growth of the photonic device layer on a sacrificial layer and selective etching of this sacrificial layer upon fabrication of the photonic devices to release them from the bulk. The ability to grow silicon on silicon dioxide in a scalable way has accelerated the field of large-scale silicon photonics \cite{ye2013review}. Similar efforts have been made to grow 3C-SiC on silicon \cite{anzalone2011advanced, bosi2013optimization, lu2014high}, but such a SiC layer develops defects up to thickness of a few microns due to the large difference in the lattice constants of the two materials \cite{calusine2014silicon}. This has encouraged researchers to develop unconventional methods such as the Smart-Cut method \cite{yi2020wafer}, photoelectrochemical etch \cite{bracher2017selective}, and wafer bonding and thinning \cite{lukin20204h}. Each of these methods has limitations in achieving scalability, such as implantation damages (Smart-Cut method), requirement of specific doping profiles (photoelectrochemical etch), and uniformity over wafer-scale (wafer bonding and thinning). 

As color centers are best defined in bulk SiC, a bulk processing method is required to unlock scalability. Angle etching is one such bulk processing technique that fabricates suspended photonic devices; it was first demonstrated in diamond \cite{burek2012free} and later in SiC \cite{song2018high, babin2022fabrication}. These angle etching demonstrations were Faraday cage-assisted, in which the ions in the reactive ion etch chamber are directed at an angle by the Faraday cage to etch underneath the mask, producing suspended photonic devices with triangular cross-sections. Finite-difference time-domain (FDTD) modeling of triangular cross-section photonics has shown suitability for color center integration and QIP hardware applications \cite{majety2021quantum, babin2022fabrication, majety2023triangular, majety2023beamsplitter, saha2023utilizing}. However, Faraday cage-assisted angle etching is not scalable beyond the millimeter chip scale \cite{latawiec2016faraday} and requires maintaining an inventory of Faraday cage designs for fabricating photonic chips designed for different applications and color center wavelengths. 

As an alternative to the Faraday cage method, Ion Beam etching (IBE) was used to achieve angle etching on a chip-scale in bulk diamond \cite{atikian2017freestanding}. In this method, ions from an ion source are accelerated towards the substrate, and the substrate holder can be tilted to etch underneath the mask to create an undercut. Color centers integrated into triangular photonic devices in diamond fabricated using IBE were used to make several key QIP demonstrations such as nanocavities with cooperativities exceeding 100 \cite{bhaskar2020experimental}, $>$99\% single-shot readout and spin initialization fidelities \cite{bhaskar2020experimental}, cavity-coupled two-qubit registers with seconds-long coherence times \cite{PhysRevLett.123.183602, PhysRevB.100.165428, doi:10.1126/science.add9771}, generation of single photons with highly tunable temporal wave packets and high spectral purity useful for generation of cluster states \cite{PhysRevLett.129.053603}, $>$ 90\% outcoupling efficiencies from emitter-cavity system into fiber \cite{bhaskar2020experimental, 10.1063/5.0170324}, memory enhanced quantum communication \cite{bhaskar2020experimental}, strain control for emitter tuning \cite{PhysRevX.9.031022, 10.1063/5.0171558}, and quantum frequency conversion to the telecom band with high indistinguishability \cite{PRXQuantum.5.010303}. As high purity 4H-SiC wafers are commercially available and with several key QIP demonstrations missing, developing a scalable angle-etch process in 4H-SiC using IBE is imperative. 


In this work, we examine a novel method for the fabrication of suspended photonic devices with triangular cross-sections on a wafer-scale in bulk 4H-SiC using reactive ion beam etching (RIBE), a variation of IBE. We develop a RIBE process to etch a variety of suspended photonic structures for a range of angles. Next, we demonstrate the wafer-scale nature of the etch process for a 5-inch diameter wafer, which is the first such wafer-scale process demonstrated for etching suspended structures in any bulk material. Finally, we use photoluminesence measurements to confirm that the optical properties of the color centers integrated into the fabricated photonic structures were preserved, which is key to building QIP hardware with color centers as quantum emitters.    
\section{Wafer-scale etching of 4H-SiC}
Ion beam etching is a physical milling process in which inert gases such as Argon are ionized in an RF ion source, and these ions are extracted as a collimated beam through a series of grids and accelerated towards the substrate to etch the target substrate. Variations of this process, reactive ion beam etching (RIBE) and chemically assisted ion beam etching (CAIBE) have a chemical etch component and are particularly useful for selective etching or etching materials that are tolerant to milling. In RIBE, inert gases in the ion source are substituted by reactive gases (SF$_6$ and O$_2$ for SiC), and the reactive gas ion beam accelerated towards the target substrate generates the etch. On the other hand, in CAIBE, inert gases are used in the ion source and reactive gases are separately introduced into a gas ring close to the target substrate. The accelerating inert gas ion beam, upon striking the reactive gas molecules close to the target substrate produces reactive gas ions that generate the etch. While both of these variations generates desirable etch outcomes, RIBE offers superior etch mask selectivity and wafer-scale uniformity \cite{shul2011handbook}, whereas CAIBE requires low maintainance for the ion source \cite{kaufman2011applications}.

\begin{figure}[htbp]
    \centering
    \includegraphics[width=\textwidth]{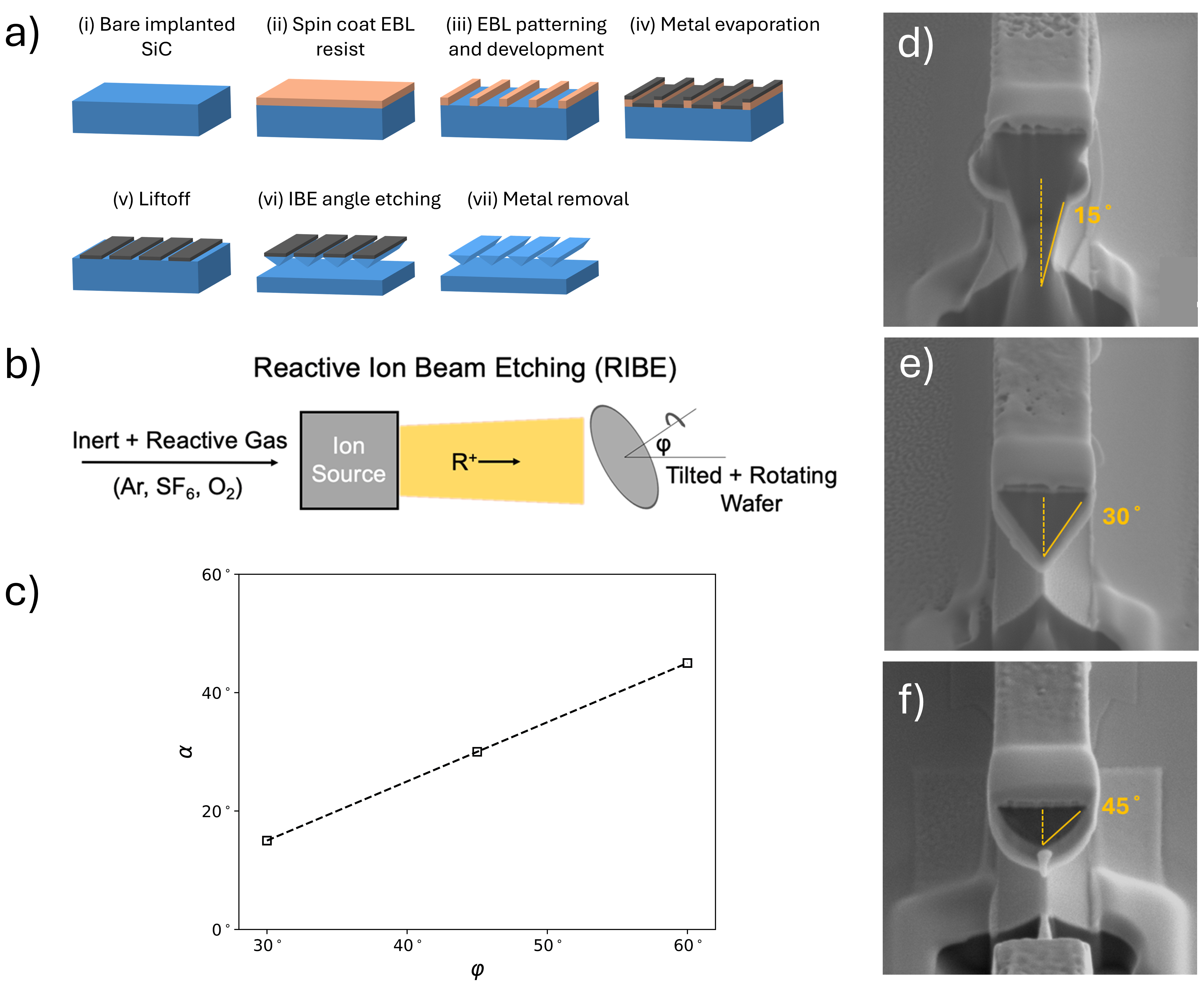}
    \caption{a) The process flow for the nanofabrication of triangular cross-section photonics in 4H-SiC. b) Schematic of the Reactive Ion Beam Etching (RIBE) with a tilting substrate holder, used to nanofabricate suspended structures with triangular cross-section in bulk 4H-SiC. c) Variation of the etch-angle of triangular structures ($\alpha$) as a function of the tilt angle of the substrate holder ($\varphi$). d-f) SEM images of the triangular cross-section profile of a 1 $\mu$m wide waveguide, showing etch angles for tilt angles d) 30$^{\circ}$, e) 45$^{\circ}$, and f) 60$^{\circ}$. }
    \label{fig:fig_1}
\end{figure}

The triangular cross-section photonic devices were nanofabricated using the process flow as shown in Figure \ref{fig:fig_1}a (details in \nameref{sec:Methods}). We developed a RIBE process involving SF$_6$ and O$_2$ gas chemistry, while maintaining the SF$_6$ to O$_2$ flowrates at a ratio 4:1 (etch process parameters in \nameref{sec:Methods}). The undercut in these photonic structures was achieved by tilting the substrate holder (Figure \ref{fig:fig_1}b), to direct the ions underneath the nickel hard mask layer and rotation of substrate holder was used to ensure uniform etch in all directions. The etch angle ($\alpha$), defined as the half angle at the apex of the triangular cross-section, varies linearly with the tilt angle of the substrate holder ($\varphi$), as shown in Figure \ref{fig:fig_1}c. The realized etch angle of the fabricated devices deviates from the expected etch angle, determined by the tilt angle, by 15$^{\circ}$ due to secondary fabrication processes. Figure \ref{fig:fig_1}d-f shows the SEM images of the cross-section of a 1 $\mu$m waveguides etched at tilt angles 30$^{\circ}$, 45$^{\circ}$, and 60$^{\circ}$ respectively. The devices etched at a tilt angle of 60$^{\circ}$ (Figure \ref{fig:fig_1}f) is the highest etch angle (45$^{\circ}$) reported for triangular cross-section devices fabricated in 4H-SiC, with an undercut of $>$ 1$\mu$m.  

\begin{figure}[hb]
    \centering
    \includegraphics[width=\textwidth]{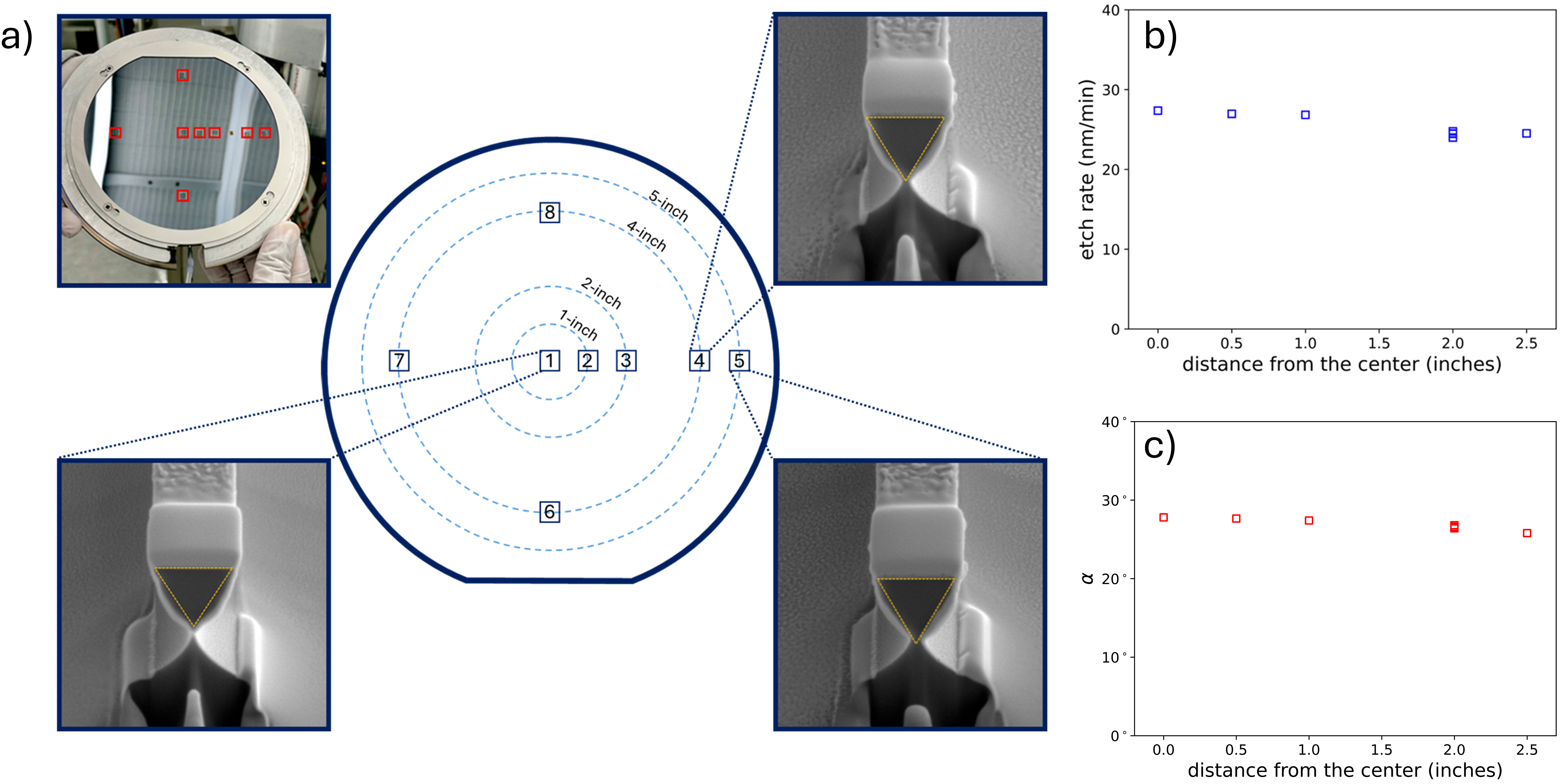}
    \caption{a) Illustration showing the placement of 4H-SiC samples on a 6-inch substrate holder. The positioning of the 4H-SiC samples represent circumferences of wafers up to 5 inches in diameter. (Insets) 4H-SiC samples on the substrate holder and the cross-sectional profiles of the nanofabricated 1 $\mu$m waveguides from some of the illustrated chips. b-c) The achieved b) etch rate and c) etch angle ($\alpha$) uniformity as a function of distance from the center of the substrate holder, showing 5.4\% and 2.9\% variability, respectively.}
    \label{fig:fig_2}
\end{figure}

We tested the wafer-scale nature of the RIBE process by placing a total of eight patterned 4H-SiC samples on a 6-inch silicon substrate holder at locations as follows: one sample each at the center, on the circumferences of 1-, 2- and 5-inch wafers, and four samples on the circumferences of a 4-inch wafer, as shown in Figure \ref{fig:fig_2}a. All the eight samples were etched simultaneously using the RIBE process developed at a tilt angle of 45$^{\circ}$ for a total of 70 minutes. The SEM images of the cross-section profiles of a 1 $\mu$m waveguide on samples placed at the center and on the circumferences of 4- and 5-inch wafers are shown in the insets of Figure \ref{fig:fig_2}a. We find that both the etch rate and the etch angle decrease linearly along the radius of the wafer, as shown in Figure \ref{fig:fig_2}b-c. The variation of etch angle along the radius could be attributed to the non-uniformity of the ion beam profile for a broad beam Kaufman-type ion source \cite{dornberg2023characterization}, across a 6-inch diameter. The radial reduction in the ion current density  and ion directionality contributes to a reduction in the etch rate and etch angle respectively. The etch rate and etch angle change linearly by 12.3\% and 5\%, respectively, from the center to the circumference of a 5-inch wafer. The variation in etch rate and etch angle among the four samples on the circumference of a 4-inch wafer was 1.2\% and 0.6\% respectively. This RIBE process has etch rate of $(25.9 \pm 1.4)$ nm/min, etch angle of $(27 \pm 0.8)^\circ$, and etch selectivity of $>$ 8:1. To our knowledge, this is the first demonstration of a wafer-scale etch process for nanofabricating suspended photonic structures in bulk 4H-SiC.


\begin{figure}[htbp]
    \centering
    \includegraphics[scale=0.8]{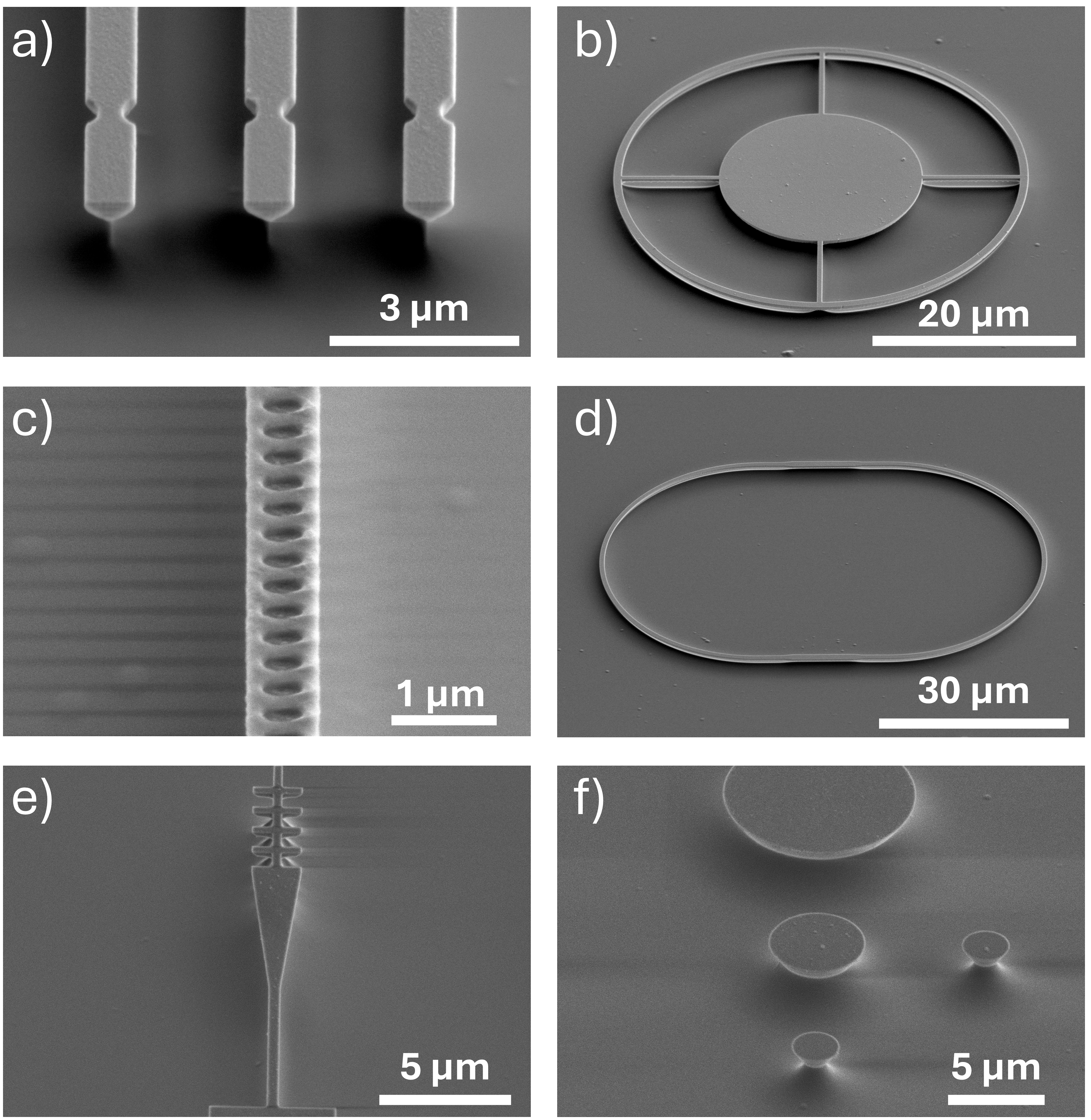}
    \caption{SEM images of the nanofabricated SiC photonic devices using the wafer-scale angle-etch RIBE process: a) 0.8 $\mu$m waveguides with notch couplers, b) microring resonator with radius 20 $\mu$m, c) photonic crystal cavity of width 0.9 $\mu$m, with 0.495 $\mu$m spacing between adjacent elliptical holes, d) racetrack resonator with a width of 0.8 $\mu$m and radius of curvature 24.6 $\mu$m, e) fish-bone grating coupler with teeth of width 2 $\mu$m and spacing of 0.8 $\mu$m, and f) microdisk resonators with radii 1.25 $\mu$m, 2.5 $\mu$m and 5 $\mu$m. SEM images in a-d) are obtained with the nickel metal hard mask on top of the devices, while e-f) have the nickel mask removed.}
    \label{fig:fig_3}
\end{figure}

\begin{figure}[htbp]
    \centering
    \includegraphics[width=\textwidth]{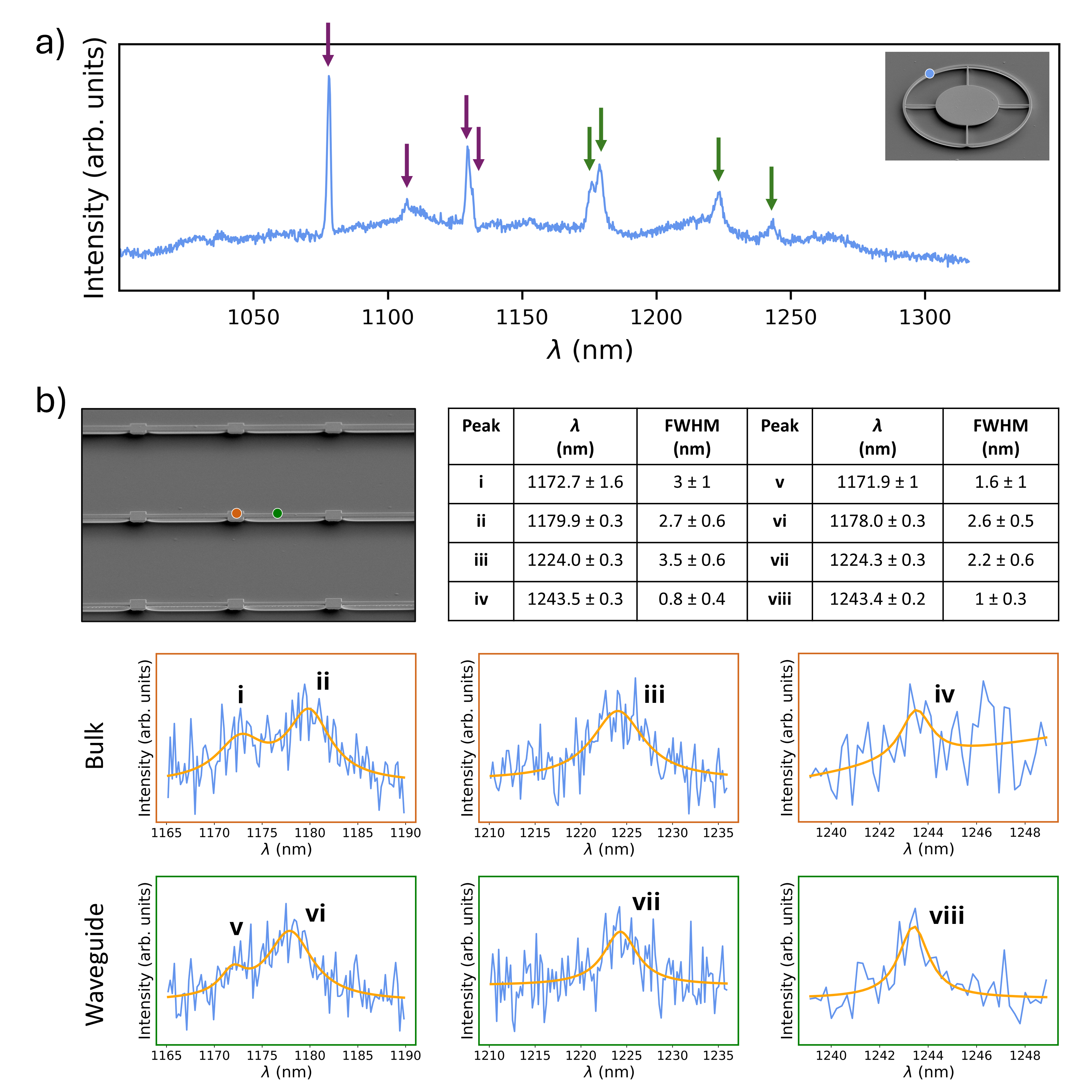}
    \caption{a) Photoluminescence spectrum of an angle-etched 4H-SiC ring resonator with integrated color centers at $T = 7$ K, showing ZPLs from divacancy ($\mathrm{VV^0}$) centers (1070 - 1130 nm) and NV centers (1170 - 1245 nm), marked with red and green arrows respectively. (Inset) SEM image of the triangular cross-section ring resonator fabricated using the optimized wafer-scale RIBE process, the blue dot indicates the spot where photoluminescence spectra were obtained from. b) High resolution comparison of NV center ZPL photoluminescence from 4H-SiC bulk (marked by an orange dot in the SEM image) and waveguide (green dot) measured at $T = 1.56$ K with peak and linewidth fits to a lorentzian curve shown in the table.}
    \label{fig:fig_4}
\end{figure}

The RIBE etch performs uniformly for a range of neighboring active and passive nanophotonic devices, as shown in Figure \ref{fig:fig_3}. We measured the emission properties of the nitrogen vacancy centers integrated in the fabricated triangular photonic devices using our 3-in-1 integrated cryogenic system for emission, collection and photon-detection \cite{norman2024icecap}. The photoluminescence spectrum of a triangular cross-section ring resonator at 7 K is shown in Figure \ref{fig:fig_4}a. The spectrum features zero-phonon lines (ZPLs) corresponding to the naturally occurring divacancy centers (red) with wavelengths 1078 nm, 1108 nm, 1131 nm, 1132 nm \cite{falk2013polytype} and the implanted nitrogen-vacancy centers (green) with wavelengths 1175 nm, 1179 nm, 1222 nm, 1243 nm \cite{mu2020coherent}. This shows that the key challenge in integrating color centers with photonic devices is solved in our process, as many fabrication processes destroy the ZPL emission \cite{calusine2014silicon}. To measure an extent of the emission preservation we take high resolution photoluminescence spectra at $T = 1.56$ K shown in Figure \ref{fig:fig_4}b where we compare NV center emission from bulk (unetched pads) and from triangular waveguides. Due to the background noise, the signal-to-noise ratio is modest, however, the inhomogeneous linewidths broadening between the two sets of spectra are quite comparable. This indicates that there is no significant strain or amorphization generated and that color center optical properties are preserved upon integration into triangular cross-section photonic devices fabricated using the developed RIBE process.

\section{Discussion}
The data in previous sections show results of what we believe is the first wafer-scale angle etch process using ion beam etching to fabricate triangular cross-section photonics in quantum-grade bulk 4H-SiC. Importantly, the etching process does not interfere with the optical properties of the photonic integrated color centers. This approach can be used to fabricate devices with a wide range of etch angles. Over a 5-inch wafer diameter, the variation in the etch rate and the etch angle is 5.4\% and 2.9\%, respectively, which can be countered by adjustments in the lithographic mask. 
In comparison, the silicon carbide on insulator (SiCOI) approach quotes a variation of 1 $\mu$m in thickness across a 4-inch wafer \cite{lukin20204h}, which could potentially triple the device thickness across the wafer if fabricated in a unified process. 

Previous modeling results have shown that variation in etch angles of triangular photonic crystal cavities shifts the resonant wavelength of the cavity without significantly affecting the quality factor and the mode volume \cite{majety2021quantum}. The change in etch angle from the center to 2.5 inches away from the center results in a 1.2\% change in the resonant wavelength, and this can be compensated by including a variation of 1.6\% in the lattice constant within the mask design to counter this resonant wavelength shift. For triangular waveguides, a variation of 2.9\% in the etch angle does not significantly change the coupling efficiencies of the color center emission or the single-mode nature of these waveguides \cite{majety2021quantum, babin2022fabrication}. The variation in the etch rate does not play a significant role in the fabrication of triangular devices because it does not affect the shape of the device, but only the height of the undercut region. The origin of the 15$^{\circ}$ deviation in the etch angle from the ideal scenario, where the etch angle is equal to the tilt angle, is yet to be fully understood. Some possible explanations for the origins of this issue could be shrinking of the metal mask at the edges during the etch or over-etching of the sidewalls and the apex of the triangular cross-section during the additional etch time required to create an undercut, which would require further investigations.

This process will have applications in the scalable fabrication of uniformly performing linear, nonlinear, and quantum photonic devices, such as waveguides, microresonators, and photonic crystal cavities, as well as in complex quantum optical circuits such as mesh photonics with color center emitters in silicon carbide. Applications in optomechanics, optofluidics, and biosensing can also be supported in this platform by taking advantage of its acoustic properties and chemical inertness.

\section{Methods} \label{sec:Methods}
Prior to the lithographic process, we generated nitrogen vacancy centers in 4H-SiC through commercial (CuttingEdge Ions, LLC) implantation using $^{14}$N$^{+}$ ions at an energy of 65 keV and a dose of 1 $\times$ 10$^{14}$ cm$^{-2}$ \cite{wang2020coherent}, resulting in a peak nitrogen concentration at a depth of $\sim$ 100 nm, as calculated by stopping and range of ions in matter (SRIM) simulations. This was followed by an annealing step in a 1-inch Lindberg Blue tube furnace at 1050 $^{\circ}$C in nitrogen atmosphere for 60 minutes \cite{wang2020coherent}, to activate the implants. The process flow used for nanofabrication of triangular photonic devices in bulk 4H-SiC is shown in Figure \ref{fig:fig_1}a. First the 4H-SiC samples were spin coated with a 350 nm electron beam resist PMMA layer, and photonic patterns were transferred into this layer using a 100 keV electron-beam lithography (ii-iii). Then e-beam evaporation was used to deposit a 5 nm titanium adhesion layer and a 120 nm nickel hard mask layer, and lift-off was used to transfer the photonic patterns to this metal layer (iv-v). Triangular cross-section photonic devices were etched in 4H-SiC using the Intlvac Nanoquest II ion beam etching tool and finally the metal layers were removed by wet etching the samples in Transene's nickel and titanium etchants (v-vi). 

The SiC chips with Ni patterns were etched using RIBE in the Intlvac Nanoquest II, equipped with a 16 cm diameter Kaufman \& Robinson RF ion source and a tilting and rotating stage. The optimized RIBE process parameters are: beam voltage = 300 V, accelerating voltage = 120 V, beam current = 110 mA, RF power = 150 W, Ar/SF$_6$/O$_2$ flow rates = 10/10/2.5 sccm, pressure = 0.24 mTorr, substrate temperature = 20 $^{\circ}$C, etch time = 2 etch cycles of 35 minutes each with 2 minutes cooling time between the two etch cycles. Argon gas is added to the ion source along with SF$_6$ and O$_2$ to maintain the stability of beam parameters. 

\section{Acknowledgments}
The authors are thankful for the valuable discussions with Marko Loncar, Bart Machielse and Haig Atikian, and to Nathan Gonzalez for assisting with photoluminescence measurements. We acknowledge support from NSF CAREER (Award 2047564) and AFOSR Young Investigator Program (Award FA9550-23-1-0266). Work at the Molecular Foundry was supported by the Office of Science, Office of Basic Energy Sciences, of the U.S. Department of Energy under Contract No. DE-AC02-05CH11231. Part of this study was carried out at the UC Davis Center for Nano and Micro Manufacturing (CNM2). The authors would like to acknowledge the valuable support provided by the CNM2 staff members Ryan Anderson and Vishal Narang. 

\section{Data availability}
The datasets used and/or analysed during the current study available from the corresponding author on reasonable request.

\section{Contributions}
S.M. developed the fabrication process, S.D. performed electron beam lithography, S.M. and P.S. performed etching and scanning electron microscopy, V.N. and A.R. performed cryogenic spectroscopy, M.R. conceived the research idea and supervised the project. S.M., P.S. and M.R. wrote the main manuscript text. All authors reviewed the manuscript.

\section{Competing Interests}
All authors declare no financial or non-financial competing interests. 

\bibliographystyle{unsrt}
\bibliography{References.bib}
\end{document}